\documentclass[smus]{snow2e}

\usepackage{epsfig}
\def\beq{\begin{equation}}
\def\eeq{\end{equation}}

\def\beqn{\begin{eqnarray}}
\def\eeqn{\end{eqnarray}}
\def\alpmzsq{$\alpha_s(M_Z^2)\;$}
\def\pp{p$\overline{\rm p}$}

\begin{document}

\title{QCD: Challenges for the Future
\thanks{The work of  S.D.   supported by 
DOE contract number DU-AC02-76CH00016. The work of L.O.
supported by DOE contract number
 DE-FG02-91-ER40685. The work of W.S. supported
by DOE contract number DE-FG02-95-ER40896.}  
 }
\author{P.~Burrows$^a$,S. Dawson$^b$, L.~Orr$^c$,  and
W.~H.~Smith$^d$
\\
$^a$ {\it  MIT, Cambridge, MA~~02138}\\
$^b$  {\it Brookhaven National Laboratory, Upton, NY  11973}\\
$^c$ {\it University of Rochester, Rochester, NY~~ 14627}\\
$^d$ {\it University of Wisconsin, Madison, WI~~53705}}
\maketitle  


\thispagestyle{empty}\pagestyle{empty}

\begin{abstract}
Despite many experimental verifications of the correctness of
our basic understanding of  QCD, there
remain numerous
 open questions in strong interaction physics and we 
focus on the role of future colliders in  addressing these questions.
We discuss possible  advances in the measurement of
$\alpha_s$, in the study  of parton distribution functions,
and in the understanding of low $x$ physics at
present colliders and  potential
new facilities.  We also touch briefly
on the role of spin physics in advancing our understanding of QCD.  
\end{abstract}

\section{Introduction} 
  QCD is a successful theory of the strong interactions which
has been tested by confronting theory and experiment in both
the perturbative and non-perturbative regimes. 
It  is an unbroken symmetry with a single coupling constant, $\alpha_s$,
whose measurement over a wide range of energy scales is crucial
for verifying the consistency of the theory.  In fact, QCD is the
only theory where relativistic quantum field theory can be tested 
beyond a few orders in perturbation theory.    

Today, we have numerous measurements where the predictions of
QCD are rigorously tested.  New and increasingly more precise
jet data are being gathered in $e^+e^-$, hadron, and $ep$ collisions
to be compared with theoretical calculations beyond the lowest order.
For example, the latest results from the jet $E_T$ spectrum
obtained by CDF and D0 challenge our understanding of both
QCD jet calculations and our knowledge of
parton distribution functions.  
Before we can interpret results such as these as indications
of new physics, we must have a solid understanding of what we
expect from QCD. 
Hadron collider, photoproduction, and deep inelastic scattering data
are giving us new information on the parton structure functions, while   
recent data from HERA on the rise of the structure function, $F_2$, 
at low $x$ are stimulating new theoretical understanding.  
The study of QCD is thus a perfect example of the necessary 
synergism between theory and experiment.  
  
The discovery of the top quark provides a new arena for
testing the predictions of QCD.  Indeed, the jet energy
calibration remains the dominant source of experimental
 error, while perturbative QCD calculations stubbornly persist
in predicting a cross section slightly above the experimental
measurement.  The physics
involving the top quark has many interesting
QCD issues, but  is not covered here.
Instead it is discussed in a separate 
 contribution by the Top Quark Working Group.\cite{top}.

Despite many successes, 
however, the  regimes  of the strong interactions
where perturbative QCD is not applicable   are 
in general  
not well understood.  We have,
for example,  an incomplete understanding of quark confinement, of the
high temperature and high density phases of QCD, of the absence or unnatural
smallness of strong CP violation, and of
 how a partonic picture of QCD is connected with non-
perturbative regimes, including bound states.  The list goes on.   
  
Some of these  questions will be addressed by lattice gauge  
theory computations in coming years as more computing power
becomes available and progress is made in the development of
algorithms.  Already lattice calculations of $\alpha_s$ are 
competitive in precision
 with experimental derivations.\cite{alphas}
Reliable and convincing calculations of the hadron mass spectrum
and weak matrix elements are realistic goals for the near future.   
 
The measurement of $\alpha_s$ is a precise test of the predictions
of perturbative QCD.  Current measurements  at
the $Z$ mass are at the $\pm 5\%$
level, while the goal is a measurement of $\delta\alpha_s/\alpha_s
\sim 1\%$ over a wide range of $Q^2$.  Such measurements have the potential
to limit new physics scenarios and to constrain physics at
the GUT scale.  Section II describes techniques for measuring
$\alpha_s$ through $\nu$ physics at low $Q^2$, in $e^+e^-$ 
and $ep$ colliders at $Q^2\leq (.5-1~TeV)^2$, and at high energy
hadron colliders such as the LHC which will probe $Q^2\leq (4~TeV)^2$. 

The understanding of parton distribution functions is
critical to many measurements at present and future
colliders.  Section III contains a survey of the $x$ and
$Q^2$ regions  where progress in our knowledge of structure
functions might be obtained at future facilities.
  Particular attention is paid 
to obtaining a consistent definition of the errors in the
structure functions.
This section also discusses some problems in jet physics.  

The prospects for low $x$  and diffractive
physics are presented in Section IV.  The
scattering 
region with $x<10^{-5}$ is a new strong interaction frontier
 which has begun to be probed by HERA.  Extension of these studies
may yield insight into the role of the BFKL pomeron.  Forward
physics ($.1<x_F<.9$) and diffractive scattering can also provide
information about  the non-perturbative regime.  Possible collider 
experiments which are sensitive to forward
and diffractive physics 
 at both the Tevatron and the LHC are discussed.  
 
Spin physics has the potential to open a new window on QCD
and is discussed in Section V.  Experiments with
polarized protons and electrons can lead to new measurements
of polarized structure functions and asymmetries.  

\section{Measurements of $\alpha_s$}

Since QCD contains in principle only one free parameter, the
strong interaction scale $\Lambda$, tests of the theory
can be quantified in
terms of comparison of measurements of $\Lambda$ in different processes
and at different hard scales $Q$. In practice most QCD calculations of
observables are performed using finite-order perturbation theory, and
calculations beyond leading order depend on the
{\it renormalisation scheme} employed, implying a scheme-dependent
$\Lambda$. It is conventional to work in
the modified minimal subtraction scheme ($\overline{MS}$
scheme), and to use the strong interaction scale
$\Lambda_{\overline{MS}}$ for five active quark flavors.
If one knows $\Lambda_{\overline{MS}}$
one may calculate the strong coupling $\alpha_s(Q^2)$ from
the solution of the QCD renormalisation group equation.
Because of the large data samples taken in e$^+$e$^-$ annihilation
at the Z resonance,
it has become conventional to use as a yardstick \alpmzsq,
where $M_Z$ is the mass of the Z boson; $M_Z$ $\approx$ 91.2 GeV.
Tests of QCD can therefore be quantified in terms of the consistency
of the values of \alpmzsq measured in different experiments;
such measurements have been performed in e$^+$e$^-$ annihilation,
hadron-hadron collisions, and deep-inelastic lepton-hadron scattering
(DIS), covering a range of $Q^2$ from roughly 1 to $10^5$ GeV$^2$.

Over the past decade many measurements of \alpmzsq have been presented.
The best measurements approach a relative precision of 3\%, but 5-10\%
is typical. Within the errors all measurements are consistent with a
central value of \alpmzsq = 0.118 $\pm$ 0.005, and there is
no evidence of any discrepancy between measurements made at different
$Q^2$ values or in different processes. QCD has therefore been tested
to about the 5\%-level of precision, which is rather modest compared
with the current 0.1\%-level tests of the electroweak theory.
A primary aim of future high energy physics studies should therefore be
the achievement of much more precise QCD tests, which are necessary to
enhance our confidence in the theory, as well as to constrain
possible extensions to the Standard Model (SM) and Grand Unification
of the couplings at a high energy scale.

In this spirit the various techniques for measurement of \alpmzsq
have been reviewed, and their potential for achieving a benchmark
1\%-level of precision has been evaluated. This has involved detailed
study of the sources of uncertainty, both experimental
statistical and systematic, as well as theoretical, and projection of
the reduction in these uncertainties that may be achievable in future
experiments at existing or new facilities.

Many current measurements are limited by
theoretical uncertainties that result from truncation of the
QCD perturbation series at low order and/or from lack of knowledge of
non-perturbative effects. Only the inclusive observables $R$, $R_{\tau}$,
and the Bjorken and Gross-Llewellyn-Smith (GLS) sum rules have been
calculated perturbatively at next-to-next-to-leading order (NNLO).
Calculations of jet final states in e$^+$e$^-$ annihilation,
hadron-hadron collisions and deep-inelastic scattering are presently
limited to next-to-leading order (NLO), resulting in estimated
$\pm$5-10\%-level uncertainties on \alpmzsq due to the missing higher
order contributions. NNLO calculations of jet final states are hence a
prerequisite for improving precision beyond the 5\%-level. Though
much progress towards this goal has been made, the task is difficult
and requires further considerable theoretical effort.
It was assumed for the projection of the precisions of future \alpmzsq
measurements that these calculations will be available, and that
residual higher-order uncertainties will contribute
at or below the 1\%-level.

Uncertainties arising from non-perturbative contributions,
often called `hadronisation' or `higher-twist' effects, are
expected to have the form of a series of inverse powers
of the scale $Q$; they are hence potentially most important for
$\alpha_s$ measurements made at low scales, such as from some
structure function determinations in deep-inelastic scattering.
Though lattice gauge theory provides a successful tool for
performing non-perturbative QCD calculations, it
is currently limited in applicability to static properties of hadrons
and cannot be used to calculate
power-law corrections to hadronic final-state observables.
Instead these are usually estimated using {\it ad hoc}
parameterizations and models of hadronisation. Recently progress has
been made towards a deeper level of understanding in the form of
studies of `renormalon ambiguities', which may represent the first
step towards a `theory of hadronisation'.

Four techniques were identified that offer the best prospects for
1\%-level \alpmzsq measurements:
(1)~the $Q^2$ evolution
of the parity violating structure function $xF_3$,
(2)~the Gross-Llewellyn-Smith (GLS) sum rule,
(3)~spin-averaged splittings in the $\Upsilon$ and $\Psi$ systems,
and (4)~hadronic observables in e$^+$e$^-$ annihilations.
(1) and (2) are measured in
deep-inelastic neutrino scattering experiments;
(3) is based on lattice QCD; all other methods use
perturbative QCD.

The $Q^2$ evolution of $xF_3$ has the attractive feature that it is
independent of the gluon distribution function of the nucleon. It is
best measured using the difference between $\nu$ and $\bar{\nu}$
cross sections for scattering on unpolarized targets. A NLO
calculation is available and the NuTeV
experiment at Fermilab hopes to achieve a precision on \alpmzsq of
$\pm2.5$\% within the next few years. Further improvement towards the
1\%-level
would require a high-statistics tagged neutrino beam facility, perhaps
at the upgraded TeVatron (`TeV33') or at the LHC, and a NNLO
calculation of the DGLAP splitting functions.

The GLS sum rule has already been calculated at NNLO, but because of
the low $Q^2$-value, about 3 GeV$^2$, of present experiments,
higher-twist effects are important. The NuTeV experiment expects to
obtain a precision of $\pm$3\% on \alpmzsq using this technique.
The larger $Q^2$ values and lower $x$-reach potentially attainable with
neutrino beams at TeV33 or LHC would allow the possibility of 1\%-level
measurements.

Heavy quarkonium systems can be used to determine $\alpha_s$
by comparing the measured
energy-level splittings with a lattice QCD calculation.
The most precise determinations of \alpmzsq with this technique
to date, at the $\pm3$\%-level, have been obtained
by the FNAL/SCRI and NRQCD groups
using spin-averaged splittings in the $\Upsilon$ and $\Psi$ systems.
The precision is limited by
uncertainties relating to lattice discretization, treatment
of sea quarks, and matching between the different
renormalisation schemes used in lattice and perturbative calculations.
All of these issues can be addressed by first-principles calculation
with current computational resources, and it is expected that an
\alpmzsq determination with 1\% precision can be achieved.

The measurement of $\alpha_s$ via hadronic event shape
observables in e$^+$e$^-$ annihilation has been studied in detail
over the past decade by experiments at the CESR,
PETRA, PEP, TRISTAN, SLC and LEP colliders.
The current experimental
precision on \alpmzsq achieved by a single experiment is
at the 2-3\%-level, but all experiments are limited by the fact that
the observables are calculated only at NLO, yielding uncertainties
estimated to be at the 7\%-level from the uncalculated higher-order
contributions. NNLO calculations are required to improve this
situation. Hadronisation uncertainties at the Z energy
are at the 3\%-level,
but are expected to drop below 1\% for c.m. energies above 300 GeV,
so that an e$^+$e$^-$ collider operating in the range
$500 \leq Q \leq 1500$ GeV
would be expected to achieve a 1\%-level \alpmzsq measurement,
provided NNLO calculations were available.

In addition, two other methods, the $Q^2$-evolution of the parity
non-violating structure function $F_2$ at high $x$,
and the jet $E_T$ spectrum in high energy
proton-(anti)proton collisions,
offer the possibility to determine $\alpha_s$
with good accuracy in regions of $Q^2$ which are
complementary to those of the other measurements.
The feasibility of both of these techniques is currently the focus
of studies at HERA and the TeVatron, respectively; until the results
are known it is not possible to evaluate the potential precision of
these methods. However, it is already clear that,
to obtain sufficient events
in the kinematic region $Q^2$ $\sim$ $10^4$ GeV$^2$ and $x$ $\sim$ 0.5
for an \alpmzsq measurement at the percent level,
a HERA data sample of about 1000 pb$^{-1}$, or
a `LEP $\times$ LHC' DIS facility, would be required.
It is also clear that the TeVatron and LHC offer the greatest
lever-arm for constraining the $Q^2$-evolution of $\alpha_s$, so that
feasibility studies for these measurements are strongly encouraged.

In summary,
the goal of measuring \alpmzsq to a precision of $1\,\%$, with
a number of complementary approaches and over a wide range of
$Q^2$, seems feasible.
The determination of \alpmzsq from the hadron spectrum using lattice
QCD is the only method without facility implications.
A more complete program will likely require new facilities.
These include a tagged neutrino facility
utilizing either the full energy Tevatron beam or one of the LHC
proton beams, and a high-energy e$^+$e$^-$ collider.
The potential for complementary $\alpha_s$ determinations in
\pp$\;$ collisions at the TeVatron and pp collisions at the LHC needs
further study.

\section{Structure Functions and Jets}

QCD is a theory of quarks and gluons.  The strongly interacting particles 
observed in experiments are {\it not} quarks and gluons, but hadrons.
To make contact between theory and experiment we need to
make a connection between the partons in the theory and the particles
that are actually involved in our experiments and that show up in our
detectors.  This connection is provided by the parton distribution functions
obtained from nucleon structure function measurements and by models of jet 
fragmentation and algorithms for jet definition.  Making use of 
{\it any} physics measurement with hadrons in the initial or final state 
necessarily involves taking into account parton distributions
and/or jet physics.
In addition, many new physics signals (and `old' ones as well) have
QCD processes as major backgrounds.  
The work of this subgroup is therefore not
only integral to QCD studies, but is directly relevant
to the other physics working groups as well. 

At the 1996 Snowmass, more emphasis was placed on structure functions than 
jet physics, as will be reflected in this summary.  We will also see, 
however, that recent results from the Tevatron collider have 
closely intertwined some aspects of the two subjects.  For a 
more complete discussion the reader is referred to the Structure 
Functions Subgroup Summary\cite{sfwg} and to individual contributions
to these proceedings.

\subsection{Parton Distributions:  Where They Come From and 
Where They Might Be Going}

We understand the structure of nucleons in terms of their parton constituents.
From measurements of structure functions in such processes as deep
inelastic scattering we obtain quantitative descriptions of those 
constituents in the form of parton distributions functions (PDFs),
which describe the momentum density of the partons inside the 
proton.  Such descriptions are necessary not only to interpret
experiments in the context of QCD (and thereby test 
the theory), but also to make predictions for experiments with 
hadrons in the initial state, for example at the Tevatron and LHC colliders.

At Snowmass the PDF working group focussed on how the $x$ and $Q^2$ ranges 
might be further extended, and how we might make better use of the 
measurements we already have through more detailed exploration of the 
processes by which the PDFs are obtained.  This includes preliminary 
efforts to understand, and wherever possible to quantify, the sources
of uncertainties associated with PDFs.  The latter subject is discussed in 
the next subsection.

The quark distributions are determined primarily from deep 
inelastic scattering of both charged leptons and neutrinos off of nuclei,
with additional constraints coming e.g. from the Drell-Yan  process.  
The gluon distribution is considerably less well known and is 
determined from direct photon production (through the subprocess
$qg\rightarrow q\gamma$) and also (at low $x$) from measuring the 
$Q^2$-evolution of the structure function $F_2$.  

A recent development is the increased use of hadron collider
data as input to parton distribution determinations.  In particular,
the asymmetry in the rapidity distribution of leptons from $W$ decays
has been used to constrain the difference between the $u$ and $d$ quark valence 
distributions, and the inclusive jet $E_T$ spectrum contains information about 
the gluon at large $x$.  
When used judiciously, hadron collider data have the potential to
give information about parton distributions in previously inaccessible, or at 
least not very well constrained, kinematic regimes.
We will certainly see more of this trend in the future.

Our knowledge of the PDFs has seen a marked increase in recent years,
in both precision and range in momentum fraction $x$ and scale $Q^2$,
with the advent of the HERA $ep$ collider and with the availability of
other new data. 
Figures 1--4 of Ref.\ \cite{sfwg}  summarize the experiments 
currently used in PDF determination, and show  the $x$ and $Q$ ranges 
covered by each one.  Present experiments extend down to $x$ values
of $10^{-4}$ and up to values of $Q$ approaching 100 GeV.

The process of extracting parton distributions from fits to structure 
function data 
works roughly as follows.  QCD predicts the evolution of the 
structure functions with the scale $Q^2$, but it does {\it not} predict
their absolute values.  In particular QCD does not predict (except in 
certain asymptotic regimes) the $x$ dependence.  
Therefore global fitters begin with a parameterization 
for the $x$ dependence of parton distributions at some 
low starting scale $Q_0$ --- typically on the order of a few
GeV --- and evolve them to the higher scales relevant to the 
experiments.  They then perform global fits to 
the structure function data at next-to-leading order in QCD 
to determine the values of the parameters in their starting distributions.

In principle this process is straightforward.  In practice, however, the 
relevant experiments vary widely in origin (collider vs.\ fixed-target,
lepton-hadron vs.\ hadron-hadron initial states, etc.) 
and in quantity and quality --- in particular, in the
sources and sizes of errors, which tend to be dominated by systematics.
Further complications arise on the theory side from the 
necessity of making an ansatz for the starting 
distributions as well as from issues such as renormalization-scale
dependence.  The bottom line is that, in practice, fitting PDFs 
requires making many judgement calls and 
as much  art as  science.  The people who do global fits have
known this all along, of course, but as more and better data accumulate
and new questions arise (e.g. regarding jet $E_T$ distributions; see 
below), it is becoming more clear that users of PDFs should be aware
of the subtleties involved.

Global PDF fits are performed by two groups:  MRS (Martin, Roberts, Stirling)
\cite{MRS} and CTEQ (Coordinated Theoretical-Experimental Project on
QCD) \cite{CTEQ}.  
The fits are more or less continuously updated as new data
become available.
It is important to realize that {\it MRS and CTEQ are doing 
essentially the same thing};  they are performing global fits to 
more or less the same data using the procedure outlined above.  
Differences between their parton distributions arise from differences in
judgement calls they are required to make in the process of fitting the data.
The basic differences between CTEQ and MRS --- which translate
into {\it small} differences in the partons they generate --- can be 
summarized as follows.
\begin{enumerate}
\item  Their starting parameterizations for the $x$-dependence at the scale $Q_0$
are slightly different; CTEQ uses one more parameter than MRS does.
\item CTEQ does not include CCFR neutrino DIS data at small $x$, which do 
not appear to
agree with lepton DIS data.  MRS includes the CCFR small $x$ data.
\item MRS uses a fixed renormalization scale for their fits to direct 
photon data; CTEQ allows the scale to float.
\item Adjustment of the overall normalization of the various data sets differs
slightly between MRS and CTEQ.
\item Minor details of the computation of structure functions in their
fitting routines are likely to differ somewhat.
\end{enumerate}
Thus parton distributions generated by CTEQ and MRS are very similar, but
not identical.  

The Structure Functions Subgroup considered how the kinematic range 
of PDF determinations can be extended in future experiments.  It is 
desirable to go to higher $Q^2$ and especially to lower $x$, for example
to study diffractive phenomena and study BFKL physics.  Because it is
unlikely that a dedicated structure function facility is a realistic option
for the future, 
the  working group examined the reach achievable by combining the 
various lepton and hadron beams proposed for the overall Snowmass studies.
Results are presented in Table V and Figures 21  and 23 (for $\sqrt{s}\approx 
1\ {\rm TeV}$ and $2\ {\rm TeV}$, respectively) of the Structure Functions
summary \cite{sfwg},
 with the reach of present and planned (i.e., approved)
facilities shown in Figure 22 for comparison.  These results include 
kinematic cuts that represent practical limitations on measuring 
the final states.  The bottom line is that,
at lower values of $Q^2$, $x$ values down to $10^{-7}$ are in principle
achievable, and for $x$ closer to $1$, the high-$Q^2$ regions can be 
better filled in.  In general, for a given $\sqrt{s}$, the smallest
values of $x$ are best achieved with a high energy hadron beam 
colliding with a low energy lepton beam, with only minimal loss
at high $Q^2$.

\subsection{Towards a Better Understanding of PDF Uncertainties}

Any prediction that uses parton distributions has some uncertainty 
associated with the PDFs that comes from uncertainties in the original
structure function data and from the method 
used to fit the data.  
Obviously we would like to be able to 
quantify that uncertainty.  As our high energy physics measurements 
become more and more precise, this issue becomes more and more 
important, and can mean the difference between discovering new physics
and recognizing `old' physics for what it is.  Or worse,
missing out on some new physics signal altogether.
Therefore a great deal of effort was expended at Snowmass by the 
Structure Functions working group to attempt to better understand,
and eventually to quantify, PDF uncertainties.

It has been common to estimate the contribution of these
uncertainties by performing the same calculation using different sets
of parton distributions and identifying the resulting variation with the 
PDF error.  It should be obvious from the previous subsection
that this is {\it not} the appropriate thing to do.  
MRS and CTEQ use the same data in their fits, and 
uncertainties in these data are not reflected directly in the fits.
A CTEQ-MRS comparison gives,
at best, an estimate of the uncertainty due to the differences in their
procedures outlined above; it does not give anything remotely resembling the 
sort of $\pm 1\sigma$ errors we would like to be able to compare with 
experimental measurements.  This becomes more true as new data 
 constrain the partons ever more tightly.
This is illustrated in a study \cite{sfwg}
 of $W$ mass and asymmetry measurements
at the Tevatron.  The $W$ mass measurement 
has improved to the point where the associated PDF uncertainty is 
becoming very important.  An attempt to bracket the PDF errors simply
by varying PDF sets is not sufficient; see Section IV of \cite{sfwg}
 for details.

What then {\it is} the appropriate thing to do to estimate PDF uncertainties?
There is no clear answer yet.
Some of the difficulties and potential pitfalls involved in determining 
PDF errors are already implied above.  The Structure Functions 
subgroup identified a number of such issues that deserve further study
which we summarize here; see \cite{sfwg} for details of their studies.

\subsubsection{Sources of uncertainties}

(1) {\it Data used in the fits.\/}
 
If all of the data were independent and all sources of uncertainty 
statistical, a straightforward $\chi^2$ analysis would be valid and the 
results could be taken at face value.  But in the real world the data
that go into PDFs come from widely varying experiments and are dominated
by systematic errors, and the sources of these systematic errors 
also vary widely because of the differences in physical processes involved.
For example, jet energy uncertainties may dominate one experiment, while
another is complicated by nuclear effects, for which corrections must
be applied, and so on.  In addition, some measurements are more
difficult or complicated than others, and some systematic
uncertainties correspondingly more difficult to estimate.  While in
some cases systematic errors can be estimated quite reliably, in 
others the best we can do is an educated guess.

To confound matters further still, the errors within (and sometimes between)
experiments are often correlated.  A proper treatment would require
taking this into account, which would in turn require having detailed 
information about the correlations, e.g.\ in the form of correlation
matrices.  Such information has not been available until recently, when 
several experimental groups have begun providing it\cite{CORR}.

(2) {\it Theory.\/}
  
There are uncertainties on the theory side as well, stemming mostly
from the fact that we can only calculate to finite order in perturbation
theory.  This is manifest in things like renormalization scale
dependence and higher twist corrections.  There are also questions such as
how to correct for nuclear effects and how to handle the charm mass
in DIS.  It is tricky to estimate uncertainties associated with 
these issues; the very reason we need to estimate them in the 
first place --- viz., that we don't know the exact results --- also
guarantees that we cannot know how large an error we are really making.
In general our best bet is to perform calculations to the highest
order manageable, thereby minimizing the errors themselves.
For estimating those errors, we must use sophisticated guesses, 
for example from varying the renormalization and factorization scales.
In addition, some of these errors are also correlated, and even if we could
estimate them reliably, it is far from obvious how to set up a theoretical
error matrix.  

(3) {\it The fitting procedure itself.\/}
   
In addition to those with the data and the theory,
the fitting procedure itself introduces some uncertainties.  These
involve the judgement calls indicated above.  For example, what
should be done when two sets of data disagree outright, or if a good fit 
to one set comes only at the expense of a good fit to another?  
What constraints should be applied to the data, e.g. sum
rules from the theory, or precision measurements (such as that of 
$\alpha_s$) from other experiments?  How should we choose a starting 
parameterization for the parton distributions in the absence of 
guidance from the theory?  Is it better to have more parameters 
for more flexibility, or fewer parameters for less arbitrariness?
Should the experimentalists' (and theorists') estimation of their 
errors be taken at face value?  Each of the myriad choices required in 
performing global fits to such disparate experiments is a potential
source of uncertainty.  How to quantify those uncertainties is at this point
as subjective as the choices themselves.

In fact a preliminary study has been done for one of these issues; see 
section IIE of \cite{sfwg}.  The
uncertainties associated with differences in parameterization can be quantified,
at least in the context of the particular parameterizations used by
MRS and CTEQ.  The purpose was to investigate how well CTEQ partons could
be modeled by MRS.  A fit was performed to CTEQ3M partons at the 
starting scale $Q_0$ using the more restrictive MRS 
parameterization.\footnote{Note that the difference is largest at the 
starting scale $Q_0$ because QCD evolution washes out differences 
as $Q^2$ increases.}  The study compared the MRS-parameterized fit to 
the original CTEQ fit as a function of $x$ for the gluon, $u$ valence,
and $d$ valence distributions.  The deviation between the two sets
is never more than 2\%, and is less than 1\% for most of the $x$ range,
the exception being for values of $x$ approaching 1.  This preliminary
study shows that the difference in parameterizations is relatively
minor source of differences between MRS and CTEQ, and
presumably the lack of sensitivity of the fits to the 
details of the parameterization also shows that it is
not a significant source of uncertainty in the fits themselves.

\subsubsection{A realistic goal:  `custom' parton distributions}

The preceeding discussion suggests that it is very difficult --- and 
possibly even meaningless --- to attempt to distill the disparate
sources of uncertainty into a single, all-purpose PDF error.
It is perhaps more sensible, and certainly more realistic, to 
tailor the estimation of PDF uncertainties to a given physical process.
That is,  we can determine what particular distributions in
what $x$ range dominate in the specific physical
process of interest.  Then global fits to parton distributions can be generated
by varying within their allowed ranges the data sets that contribute
in the relevant region, while still requiring a good fit to the 
remaining data.  In fact such an approach has already been taken 
for distributions relevant to inclusive jet $E_T$ distributions at the 
Tevatron.  In \cite{MRSALPHA}, the MRS collaboration generated PDFs with
different values of $\alpha_s$ allowed by various data sets, and the CTEQ 
collaboration has generated a set of PDFs (CTEQHJ \cite{CTEQHJ}) which 
are tailored specifically to accommodate the high $E_T$ jet data.
Another possible application is to the $W$ mass measurement, which is
sensitive to the $u$ and $d$ quark valence distributions; one can vary 
them to the extent permitted by the $W$ asymmetry data to get a first
shot at estimating the PDF uncertainty.

This approach was strongly advocated at Snowmass \cite{sfwg}.
It requires a bit more sophistication on the part of PDF users (not to 
mention more work for the people who perform the global fits!),
but for the forseeable future it seems to be our best bet for obtaining 
reasonable estimates of PDF uncertainties.

\subsection{Case in Point:  Jet $E_T$ Distributions}

\subsubsection{High $E_T$}

Many of the points mentioned above are nicely illustrated by the study
of the jet $E_T$ distributions recently measured by CDF and D$\emptyset$
\cite{CDFJ,DZEROJ}.
The inclusive jet $E_T$ distribution has long been held up as evidence
for how well QCD works; the agreement between data and NLO QCD as the 
cross section falls by at least six orders of magnitude is indeed
impressive.  But recently CDF measured an excess above NLO QCD expectations
\cite{NLOQCD} for 
$E_T > 200~{\rm GeV}$ \cite{CDFJ}.
Is this new physics, or is there
simply something missing in our comparison?  It is our duty to rule 
out every possible Standard Model explanation before concluding, e.g.,
that we have observed compositeness.  This is particularly true 
because D$\emptyset$ does not see a clear excess at high $E_T$ \cite{DZEROJ}; 
however, their results are not inconsistent with CDF's, because their 
(D$\emptyset$'s) errors are larger.

A number of questions then arise as to how well we really know the uncertainties
in the theory and the experiment.  On the experimental side, how well
is jet energy measured at high $E_T$?  Are uncertainties associated with 
such factors as the
 jet algorithm and the various necessary corrections 
    under control?
On the theoretical side, how large
can we expect higher-order corrections to be?  What about renormalization-
and factorization-scale dependence?\footnote{A recent paper suggests \cite{KK}
that the discrepancy may be accounted for by factorization {\it scheme}
dependence.}  What is the uncertainty in the QCD prediction due to 
parton distributions --- is there enough leeway there to bring theory
and experiment back into agreement?  

The latter question was of considerable interest to this working group.
MRS have shown \cite{GMRS} that it is not possible to bring the CDF 
results into agreement with the theory by adjusting the quark distributions
without spoiling the global fit to other data.  The CTEQ collaboration
focussed on the gluon \cite{CTEQHJ} and showed that, with an additional
parameter, it is possible to adjust the gluon distribution at large
$x$ to accommodate the CDF data, because the gluon is not so 
well constrained in that region.  

There are several points to be made here.
First, this exercise shows that comparing PDF sets does not give a 
realistic reckoning of how much variation is possible in the PDFs.
A comparison of MRS and CTEQ gluons in this region would give a difference
on the order of 10--20\%.  But the CTEQHJ gluon wound up larger by as much 
as a factor of
two than previous ones, and that was without a significant sacrifice in the 
quality of the fit.  Second, we see that some distributions are more
tightly constrained by existing data than others.  Third, we want our
parameterizations to be flexible but not arbitrary.  Fourth, this 
provides an example of the `custom' parton distributions described above.
Finally, we will soon have independent information about the gluon
in this region from the E706 prompt photon experiment. 

\subsubsection{Medium $E_T$}

\begin{figure}[h]
\epsfig{file=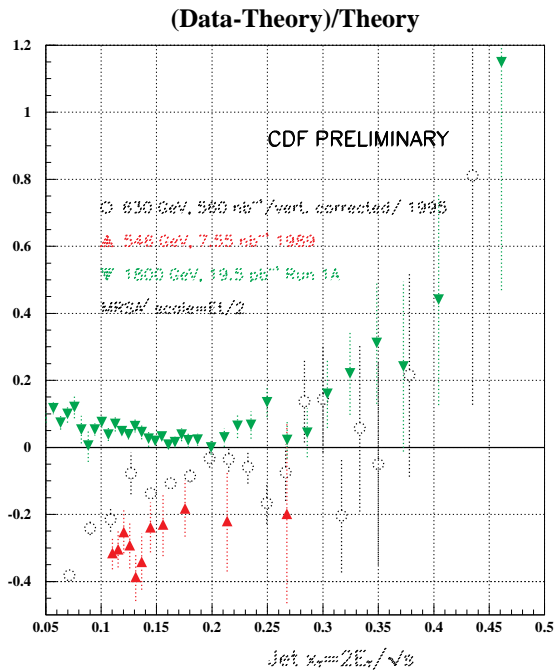,height=3.5in}
\caption{Preliminary CDF measurement of inclusive jet $x_T$ distribution
at center-of-mass energies 546, 630, and 1800 GeV \protect\cite{CDFNEW}. }
\label{jetfig}
\end{figure}

The bottom line of the above exercise is that the high-$E_T$ jet data 
{\it can} be brought into agreement with the theory by exploiting 
the flexibility in the gluon distribution at large $x$.  Before
breathing a collective sigh of relief, however, we must 
come to terms with a problem that has yet to be resolved:
the medium-$E_T$ jet distribution.
In 1995 the Tevatron ran briefly at a center of mass energy of
630 GeV, and CDF measured the inclusive jet $E_T$ distribution
\cite{CDFNEW}.  If the distribution is expressed in terms of
the variable $x_T\equiv 2E_T/\sqrt{s}$, it scales:  the distribution
is independent of center of mass energy and the parton distribution dependence
cancels out.  Thus the jet $x_T$ distributions measured at 630 and 
1800 GeV should agree.  They do not, as shown in Figure \ref{jetfig}.
There is less room for possible explanations here than above; in 
particular, the disagreement {\it cannot} be fixed with partons.  
It may be that we do not understand QCD as well as we think we do, or
it may be that we do not understand jet measurements as well as we think we
do.  In either case, it is clear that there is {\it something} we
still don't understand, and until we resolve the questions about medium-$E_T$
jets, we cannot be confident that we understand high-$E_T$ jets.

\subsection{Structure Functions and $\alpha_s$ Measurements}

The strong coupling constant $\alpha_s$
can be determined from measurements of deep inelastic structure functions,
either from their $Q^2$ evolution, or via sum rules.
The Snowmass studies focussed on what is required to improve 
these measurements to an accuracy of a few percent or better.
We summarize these issues here; for more details and how this fits in with 
other methods, see \cite{sfwg} and \cite{alphas}.
See also Section II above.

\subsubsection{$Q^2$ evolution of structure functions}

At lowest order, deep inelastic structure functions depend only on $x$ and
are independent of $Q^2$.  But QCD predicts that at order $\alpha_s$, 
the structure functions evolve with $Q^2$.  Therefore an observation
of this evolution provides a measurement of $\alpha_s$.
The best measurements in this method are obtained from non-(flavor-)singlet
structure functions because their evolution is independent of the 
gluon distribution, which, as we saw above, is not as well determined 
as those of the quarks.  Further gains can be made by using large
values of $x$, which virtually eliminates any sea quark dependence as well.
This can be done in either charged lepton or neutrino DIS.

Experimental errors in these measurements are dominated not by statistics 
but by systematics --- in particular, energy uncertainties.  Thus improved
measurements would require better calorimetry and better calibration 
techniques.  Theoretical uncertainties are dominated by renormalization 
and factorization scale dependence; these could be dealt with by computing 
the necessary higher order terms or by allowing the two scales to float and 
fitting from data.  

\subsubsection{Sum rules}

QCD predicts order $\alpha_s$ corrections to the various sum rules 
for combinations of structure functions integrated over all values of $x$.
Measuring these integrals --- the Gross-Llewellyn-Smith and Bjorken 
sum rules are used in practice --- therefore provides a complementary method
for measuring $\alpha_s$, with a complementary set of uncertainties.
The sums predicted by the theory are not subject to scale uncertainties
at the level seen in measuring $Q^2$ evolution; the difficulties here are 
mostly experimental in nature.  One has to do with low $x$:  in principle
the integration extends to $x=0$, but in practice the measurements can only 
go to finite $x$.  Therefore it is necessary to extrapolate,
which introduces uncertainties that can be difficult to estimate.  In
addition, low-$x$ measurements tend to be made at low $Q^2$, which
introduces complications associated with higher twist effects. 
But avoiding higher twist effects by going to higher $Q^2$ introduces
a new problem, namely that $\alpha_s$ is smaller at higher $Q^2$,
which reduces the size of the effect one is trying to measure 
(which is a correction to the overall sum).  This can be mitigated somewhat 
with sufficient statistics.

\subsection{Heavy Quark Hadroproduction}

Finally, some attention was devoted to heavy quark hadroproduction.
For some time measurements of heavy quark production have significantly 
exceeded theoretical predictions.  Apparently the problem was that 
fixed-order perturbative calculations were unable to account for large 
logarithms that involved the heavy quark mass and the center of mass energy
or transverse momentum.  Efforts to solve this problem by using heavy
quark fragmentation functions have led to improved agreement; recent 
efforts \cite{HEAVY} incorporating flavor-excitation and flavor-fragmentation
diagrams have improved agreement further still.  See \cite {sfwg} and 
references therein for details.

\section{Low-{\Large$\lowercase{x}$} \& Diffractive Physics}

\subsection{ep collisions}

A renewed interest in diffractive phenomena has been sparked by ep events
that occur at high $Q^2$ and small $x$. The observation at HERA of deep
inelastic scattering events with a large rapidity gap in the final state
between the proton direction and the first energy deposit in the detector is
an indication of diffractive scattering \cite{HERADF}. The flatness of the
rapidity gap distribution, as well as other properties of the events such as
independence of the cross section on $W$, are consistent with photon
diffractive dissociation off a Pomeron. Studies of these events are
providing insights into the transition from perturbative to non-perturbative
scattering and promise to provide more information as the low-$Q^2$
transition region is further mapped out.

The strong rise in the proton structure function, $F_2(x,Q^2)$ at small x
and large $Q^2$, which indicates a strong rise in the $\gamma^*p$ total
cross section, underscores the importance of understanding the role of
diffraction at low $x$\cite{diffwg}.

While previous studies of diffraction at HERA are based on the rapidity gap 
method, more recent data have been collected with Leading Proton
Spectrometers (LPS) involving ``Roman Pot Detectors". These data provide a
sample of events  with smaller statistics and different systematics but also
with a cleaner  interpretation as diffraction and with less background from
Reggeon exchanges\cite{diffwg}.

Exclusive reactions, such as elastic vector meson production, provide
stringent tests of calculations in  perturbative QCD, as well as new methods
for extracting gluon distributions.

Inclusive reactions both in deep inelastic scattering and photoproduction
have produced insights into diffractive phenomena. Rapidity gap events form
about 10\% of the total deep inelastic scattering cross section and have been
used to measure a diffractive proton structure function. Studies of hard
diffractive photoproduction at HERA have focussed on   high $p_T$ jet
production and jets separated by a large rapidity gap. These studies suggest
a dominant gluon content to the pomeron and also that production may be
taking place by a direct photoproduction in addition to resolved
photoproduction.

Additional diffraction studies at HERA with increased luminosity and
extended coverage (i.e. LPS) in the very forward proton region will
undoubtedly lead to a  better understanding of the nature of the diffractive
process and how it  relates to QCD.

Increased statistics will enable the study of the diffractive charm
structure  function which is very sensitive to the gluonic component of the
exchange  mechanism~\cite{mehta}. 

However, it is also important to consider the advantages of going to higher
CM  energies for a lepton-hadron (ie lepton-quark) collider. Studies 
suggest that in order to reach values of $x < 10^{-6}$ for $Q^2 > $2
GeV$^2$,  one should consider a high energy lepton-hadron collider option at
one of the  future hadron-hadron colliders under consideration\cite{sfwg}.  

Further exploration of color singlet exchange and searches for an
enhancement in its cross section would be enabled by an increase in the
rapidity coverage either from increased luminosity and an extended detector
coverage at HERA or  from an increase in the CM energy that would be
available at a higher energy  lepton-hadron collider.

\subsection{pp collisions}   Rapidity gaps have been found at the
Tevatron\cite{D0gap,CDFgap} and are now a subject of considerable interest.
In principle, such events are an excellent place to study high energy
semi-hard physics including the BFKL Pomeron. However, the analysis is
complicated by the presence of hadrons coming from soft interactions
involving the spectator quarks in the colliding hadrons. BFKL phenomena are
observed in pp collisions in events with either a rapidity gap between two
jets or between a jet and a beam fragment.

Single diffractive exchange occurs when one of the protons scatters almost
elastically and the other becomes a massive multiparticle
system\cite{diffwg}. Such events are used to study the structure of the
pomeron in the context of a model where the pomeron is composed of quarks
and gluons. If the quasi-elastically scattered proton is measured, the $t$
of the pomeron and its momentum fraction are known. If the quasi-elastically
scattered particle is not measured, then diffractive events are tagged by
the presence of a rapidity gap of typically more than 3 units. While there
is a higher rate of such events, their analysis requires integration over
$t$ of the pomeron and its momentum fraction.

When there are two high-$E_T$ jets in pomeron-proton collisions, it is
possible to reconstruct the momentum fractions of the scattered partons.
Both CDF and D0 have very good evidence for diffractive dijets from
observation of an excess of rapidity gaps in one beam direction. These are
single diffractive events where the high $x_F$ particle is not seen. They
correspond to about 1\% of the dijet cross section. CDF also has evidence
for diffractive  W production. With additional statistics, both experiments
should be able to constrain the pomeron structure function. Other venues for
exploration include diffractive heavy flavor production, and looking for
double pomeron exchange in events with two rapidity gaps. Both of these
processes will require substantially more statistics than presently
available. 

The increased luminosity of Tevatron Run II and  the upgrade of CDF and D0
will provide an important opportunity to enhance understanding of
diffractive physics. The principal difficulty will be the increased rate of
multiple interactions, which will tend to obscure rapidity gaps. CDF and D0
plan to substantially increase their statistics for diffractive and forward
physics during Run II. An increase in statistics of more than 2 orders of
magnitude over that acquired in Run 1c is needed to provide an adequate
study of single diffractive exchange with tagged quasi-elastically scattered
(anti)protons. This will require longer running time with a constantly
active diffractive trigger (not dedicated runs), improved acceptance, the
installation of pots on both downstream arms if possible, and improved
triggers that veto on multiple interactions. If the detectors are equipped
with pots on both arms, they will be able to study fully constrained double
pomeron events. If CDF and D0 are able to increase their rapidity coverage,
they will enhance their gap detection and also extend their very-forward gap
physics.

There is also the possibility that a new experiment might be carried out in
the C0 intersection region at the Tevatron Collider for Run II. One option
for C0 is a detector devoted to forward and full acceptance physics proposed
by the T864 group. This experiment proposes to study rapidity gaps in soft
and hard diffraction, double diffractive dissociation, the onset of BFKL
enhancement, forward strangeness, charm and beauty production, multiparticle
correlations, and forward neutrons.

Diffractive physics at the LHC promises to be a rich source of information
since certain topologies will be cleaner due to cleaner events. It will be
possible to look for diffractive Higgs events with reduced hadronic activity
in the rapidity region near the Higgs particle\cite{LHC}. Both single and
double pomeron exchange can be observed, particularly in $H \rightarrow
\gamma\gamma$ events. In addition, single diffraction at the LHC can be
studied in $b\bar{b}$ production.

The overall rapidity span at the LHC increases from that of the Tevatron by
15 to 19 units. The mass reach of diffractively produced states also
increases dramatically. An example is that for double pomeron exchange,
(with $x_F >$ 0.95) central masses extend to 90 GeV at the Tevatron and to
700 GeV at the LHC. This extended range enables the LHC to go beyond
high-$E_T$ jet physics to electroweak probes, W, Z.

A concern with rapidity gap physics at the LHC is the multiple interactions
caused by the high luminosity. The general purpose detectors, ATLAS and CMS,
also cover only about half of the rapidity range with their present designs.
A proposal that addresses these concerns is being developed for a full
acceptance detector called FELIX. It is composed of recycled components from
ALEPH and UA1 along with very forward calorimeters and trackers extending
for 450 m to enable elastic and diffractive measurements. The goal is to
measure charged particles, photons, muons and jets over the entire rapidity
range. Being a detector devoted to this physics program, it could run with
reduced luminosity to improve identification of rapidity gaps. ATLAS and CMS
could also improve their measurement of diffractive physics by installing
Roman pots to tag high-$x_F$ protons and to provide diffractive jet
triggers. These collaborations have such options under active investigation.

Beyond the LHC, the best venue for diffractive physics appears to be a very
large hadron collider with energy of 50 - 100 TeV per beam. This would yield
a rapidity range of 25 units. Pomeron-pomeron collisions of up to 5 - 10 TeV
may be reached, which puts them well into possible SUSY and Higgs sectors.
However, a machine with such beam energy will also require very high
luminosity and therefore experience as many as 100 interactions per
crossing. This suggests consideration of a second lower-luminosity
interaction region, dedicated to diffractive and forward physics, where
single interactions could be observed. One option would be to use 2 km long
partially instrumented straight sections on either side of a modest (i.e.
upgraded CDF or D0) central detector. This suggests incorporating a 4 km
straight section into future very large hadron collider designs.

\section{Spin Physics}

Polarized processes involving hadrons
 satisfy a simple generalization of the factorization theorems 
used in hadronic physics,
\beqn 
\sigma&\sim& ({\rm{Structure~ Function}})
 \times \sigma({\rm{ Hard~ Scattering}})
\nonumber \\  &&
\times ({\rm{Fragmentation~ Function}}).
\eeqn   

  The hard scattering cross sections can be calculated in
perturbative QCD, but the parton structure and fragmentation
functions must be determined experimentally.  A primary focus
of spin experiments will clearly be the measurement of 
polarized structure functions and the verification of the sum
rules relating the various structure functions.
One would also like to measure the $x$ and $Q^2$ dependences
of the various structure functions and sum rules and compare
with the predictions of NLO QCD.  

  In 1988, the
EMC $\mu N$ scattering experiment obtained a measurement of
the nucleon spin structure function that violated the Ellis
-Jaffe sum rule.
The interpretation of this violation was that either the
 strange sea in the proton is highly polarized or that the valence
quarks carry little spin, while the remainder of the spin is
carried either by the gluons or by orbital
angular momentum.  This result and the apparent violation
of the sum rule has  stimulated a variety of spin experiments.  
\subsection{Polarized Structure Functions}
Current experiments at SLAC E143 and CERN SMC have provided
measurements of the $g_1$ structure function on protons 
with $x > 4\times 10^{-3}$.
Higher energies at HERA with a polarized proton beam or
at a fixed target experiment at an NLC, could 
 allow  
for measurements down to $x\sim 6\times 10^{-3}$.
  The study of structure functions 
at low $x$ is particularly interesting since current
data show a rise in $g_1^p$ at low $x$.
  The QCD evolution equations, however, predict that
$g_1^p$ will change sign at low $x$ and higher $Q^2$
and actually become negative.\cite{sforte}
This prediction
challenges our theoretical understanding of QCD at low $x$ and
of higher twist effects which become relevant in this regime,
as well as requiring new 
 experimental data to verify the theoretical predictions.  
Ref. \cite{spinwg} discusses
the statistical accuracy which could be obtained 
at HERA or an NLC.

\begin{figure}[h]
\epsfig{file=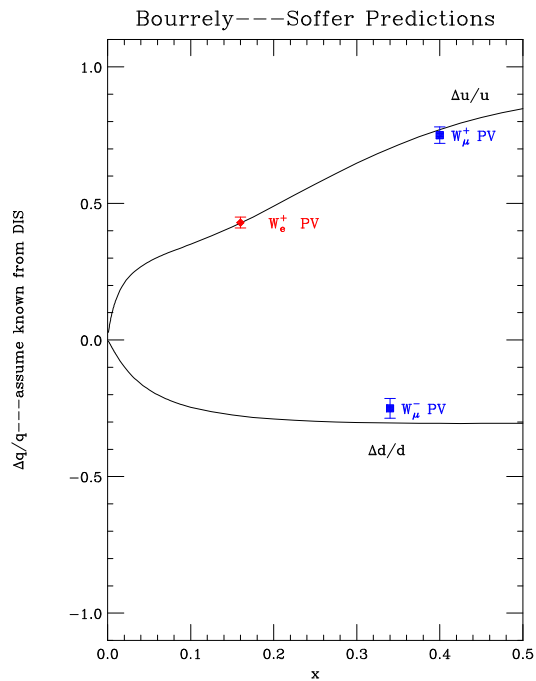 ,height=3.5in}
\caption{Expected sensitivities for $\Delta u/u$
and $\Delta d/d$ from the measurement of parity violating
effects in $W^\pm$ production at RHIC with $800/pb^{-1}$
at $\protect\sqrt{s}=500~GeV$. The solid curves are the
theoretical predictions. This figure from Ref.\protect \cite{mt}.}
\label{rhic}
\end{figure} 

The polarized beam capability proposed for RHIC offers
a unique array of spin measurements.  Both protons will be
highly polarized ($>70\%$, either transversely or longitudinally),
with high luminosity, ${\cal L}=2\times 10^{32}/cm^2/sec$, and
 energies between $\sqrt{s}=200$ and $500$ GeV.  This allows the
measurements of the gluon
structure function  $G(x)$ for nuclei, $\Delta G(x)$ for $pN$,
and  $h_1(x)$,
(which counts the valence quark polarization). 
The distribution  $h_1$ will be
measured with transverse spin asymmetries
using both $\gamma^*$  and
$Z^*$ production.  Gluon polarizations, $G(x)$ and $\Delta G(x)$, 
can be measured by using direct photons from the dominant
quark- gluon Compton scattering process, $q g \rightarrow \gamma  q$,   
and through medium $p_T$ jets, ($p_T\sim 20-50~GeV$), which are predominantly
quark- gluon produced.  $\Delta G(x)$ can then be extracted from the
longitudinal spin asymmetry, $A_{LL}$, which is predicted in NLO QCD to be
$10-20~\%$. \cite{cont}

Polarized protons at RHIC can also measure parity violating
asymmetries involving $W^\pm$ production (where the $W$ decays
leptonically).  Assuming $\Delta q$ is known from deep inelastic
scattering experiments, a precise measurement of the quark
structure  functions $\Delta u$ and $\Delta d$ can be extracted
from
the parity violating $W$ decays as shown in Fig. \ref{rhic}.\cite{mt}

\subsection{Spin Asymmetries}

Both single (one polarized beam) and double (both
beams polarized) spin asymmetries can yield useful
information about QCD.  Single spin effects in hard
scattering processes are negligible and so the measurement
of single spin asymmetries tests our understanding of higher twist
and non-perturbative physics.  The double spin asymmetries
can be used to extract moments such as
$g_1(x,Q^2)$ and $g_2(x,q^2)$, giving information
on the scale dependence and small $x$ behaviour of
the polarized structure functions.

\section{Conclusions}

There remains much to be learned from the study of QCD
and strong interactions. A precise test of the predictions
of perturbative QCD will be possible with the measurement of $\alpha_s(M_Z)$ to 
1\%. This now appears to be a realizable goal for a combination
of experiments spanning a large range of $Q^2$.  Results can be expected from
the hadronic event shapes at the NLC and also from the evolution of the
structure functions and the jet $E_T$ spectrum at future running of HERA and the
TeVatron. The
ultimate resolution would be provided by the full $fb^{-1}$ sample
planned for the HERA upgrade or a LEP $\times$ LHC DIS facility. The largest
lever arm in $Q^2$ of the planned facilities would be provided by the LHC.
We can also could expect important contributions from the GLS sum rule
measured with fixed target neutrino beams from present running with the
TeVatron and improved resolution at TeV33. Even better resolution would be
provided by a neutrino beam from the LHC or possibly from a muon collider if it
is feasible. There is also an expectation of 1\% measurements from
lattice gauge calculations of the spin-averaged splittings in the $\Upsilon$ and
$\Psi$ systems.

The measurement of structure functions and the parton density functions (PDFÕs)
determined from them are necessary for understanding high momentum processes
involving hadrons and also contain information themselves about the underlying
physics of hadrons. Understanding them is critical to understanding the
fundamental particles and their interactions. The PDFÕs are now determined over
a wide kinematic range from the HERA data at small $x$ to the TeVatron jet data
at high $Q^2$.
 There has been substantial recent progress on the determination of
the PDFÕs and their dependence on experimental data, but there is much to be
learned and considerable challenges in combining experiments with
vastly different numbers of data points and complex systematic errors.
Nevertheless, this work is very important due to the ubiquitous use of PDFÕs in
almost all experimental measurements. Examples include measurements of the W
mass and heavy quark hadroproduction. Future facilities such as LEP $\times$
LHC, a low energy lepton beam colliding with a very large hadron collider
beam, an NLC colliding with a conventional proton collider beam, or a muon
collider beam on a conventional proton collider beam would all illuminate
different regions of the $x-Q^2$ plane as shown in ref.\cite{sfwg}.
The dramatic rise of the $F_2$ with decreasing x at low x observed
in the HERA data must eventually lead to saturation of the  parton densities.
Future facilities such as these may answer where these saturation effects become
manifest. In addition, the small $x$ region can provide tests of diffractive
phenomena and resummation techniques. It appears that for fixed $\sqrt{s} $,
the best opportunities for probing small $x$ occur for a high energy hadron beam
colliding with a low energy lepton beam, in particular the preferred
$ep$ facility would be to match the highest hadron beam energy with a
modest lepton beam energy.

We are just beginning to explore and comprehend the new information on
QCD from diffractive phenomena at high energies. Additional diffractive
studies at HERA with measurement of the leading proton and higher statistics
should provide new insights into the structure of diffraction and the nature 
of the Pomeron. The increased luminosity of TeV II and the upgrade of the CDF
and D0 detectors should also enhance the understanding of diffractive physics.
If these detectors are able to increase their rapidity coverage in order to
better detect rapidity gaps, they will yield even more results. A dedicated
full acceptance experiment at C0 would provide additional opportunities.
The increased rapidity range at the LHC and the extension of the mass reach of
diffractively produced states will enable the LHC to go beyond high-$E_T$
jets physics to electroweak probes, W, Z. The multiple interactions experienced
by the general purpose LHC detectors would be alleviated a
dedicated lower luminosity full acceptance detector which could be specifically
built to have a much larger rapidity coverage. Beyond the LHC, the best
opportunity would be a very large hadron collider with an energy of 50 to 100
TeV per beam. At such a facility, pomeron-pomeron collisions are possible
with energies high enough to cover the Higgs and SUSY sectors.

Spin physics is opening a new window on QCD. Measurements of the polarized
structure functions and tests of their sum rules will provide important
information. We can expect that results from both HERA collider and NLC fixed
target data would provide a strong test of the $Q^2$ dependence of the spin
structure functions and provide information about the polarization of the
quarks and gluons. The RHIC spin program will provide complementary tests of
spin physics in a hadronic environment. Here, there is the opportunity to
search for QCD effects beyond the leading power so that the dynamics of QCD can
be studied beyond the parton model.

These investigations of QCD are important to pursue because QCD is an essential
component of particle physics. Each of the facilities considered provides
different and often complementary opportunities to investigate the wide range of
QCD phenomena. The vast range of QCD effects also underscores its importance
because it will affect almost every measurement proposed in this summer study.
This means that understanding QCD is critical to understanding the new physics
that might be observed at new facilities. The study of QCD must progress along
side the search for new phenomena and the more precise measurements of
electroweak parameters if we are to realize the full benefit of future
facilities.

\end{document}